# Adversarial Training of Denoising Diffusion Model Using Dual Discriminators for High-Fidelity Multi-Speaker TTS

**MYEONGJIN KO[1] AND YONG-HOON CHOI[1] (Member, IEEE)**

[1]School of Robotics, Kwangwoon University, Seoul 01897, South Korea

CORRESPONDING AUTHOR: Yong-Hoon Choi (e-mail: yhchoi@kw.ac.kr).

This work was supported in part by the National Research Foundation of Korea (NRF) Grant funded by the Korea Government MSIT under Grant 2021R1F1A1064080.

**ABSTRACT** The diffusion model is capable of generating high-quality data through a probabilistic approach. However, it suffers from the drawback of slow generation speed due to the requirement of a large number of time steps. To address this limitation, recent models such as denoising diffusion implicit models (DDIM) focus on generating samples without directly modeling the probability distribution, while models like denoising diffusion generative adversarial networks (GAN) combine diffusion processes with GANs. In the field of speech synthesis, a recent diffusion speech synthesis model called DiffGAN-TTS, utilizing the structure of GANs, has been introduced and demonstrates superior performance in both speech quality and generation speed. In this paper, to further enhance the performance of DiffGAN-TTS, we propose a speech synthesis model with two discriminators: a diffusion discriminator for learning the distribution of the reverse process and a spectrogram discriminator for learning the distribution of the generated data. Objective metrics such as structural similarity index measure (SSIM), mel-cepstral distortion (MCD), F0 root mean squared error (F0 RMSE), short-time objective intelligibility (STOI), perceptual evaluation of speech quality (PESQ), as well as subjective metrics like mean opinion score (MOS), are used to evaluate the performance of the proposed model. The evaluation results show that the proposed model outperforms recent state-of-the-art models such as FastSpeech2 and DiffGAN-TTS in various metrics. Our implementation and audio samples are located on GitHub.

**INDEX TERMS** denoising diffusion model, generative adversarial network, mel-spectrogram discriminator, speech synthesis, text-to-speech

## I. INTRODUCTION

Generative models [1-5] are artificial intelligence models that can generate data of types that do not exist in the training dataset. They are rapidly advancing and being applied in various fields such as natural language processing, conversation systems, art and creativity, image generation and editing, speech synthesis and conversion. The variational autoencoder (VAE) [1], which is one of the earliest and representative generative models, is a model that transforms input data into a latent space and then reconstructs it back into data. The VAE has gained significant attention and remains a prominent model in the field. On the other hand, the generative adversarial network (GAN) [2], which is still the most widely recognized generative model, generates high-quality data that is difficult to distinguish from real data using the adversarial training approach involving a discriminator and a generator. Another type of generative model, the flow-based generative model [3], explicitly learns the probability distribution of the latent space and uses inverse transformations to generate data. However, these representative models still have limitations such as unstable training, limited diversity of generated results due to mode collapse, and dependence of loss functions on the model architecture.

The diffusion model [4] is a model inspired by non-equilibrium thermodynamics used in physics. It follows the concept of gradually reducing noise while preserving important details by adding and removing noise in the data, based on the Markov chain rule where the value of each variable depends on the previous time step. This diffusion model learns the data restoration process by a fixed training procedure and can generate data from random noise or the latent space. Unlike traditional generative models, the diffusion model exhibits high training stability and does not compress data, allowing the latent space to have the same high dimensionality as the original data. In the field of computer vision, many models have been developed based on the characteristics of the diffusion model, starting with denoising diffusion probabilistic models (DDPM) [5].

A speech synthesis system, also known as a text-to-speech (TTS) system, is a system that converts a user-inputted script into speech or transforms a given reference audio into speech with different characteristics. In the early stages, TTS systems employed concatenative synthesis, specifically the unit selection synthesis [6] method, which joins phonemes together, as well as statistical parametric speech synthesis techniques such as hidden Markov models (HMM) [7]. However, these methods had limitations in generating natural-sounding synthesized speech. With the advent of neural network models, speech synthesis models have rapidly evolved.

One prominent neural network-based speech synthesis model is Tacotron [8]. Tacotron includes an acoustic model that applies the sequence-to-sequence architecture [9] and attention mechanism [10] to generate mel-spectrograms in an autoregressive manner. These mel-spectrograms are then transformed into linear spectrograms, and the audio waveform is generated using the Griffin-Lim vocoder [11]. Tacotron2 [12] improves upon the original Tacotron's acoustic model and adopts WaveNet [13] vocoder instead of [11]. While Tacotron2 addresses the issues of unnatural speech and complex hyperparameter tuning that were present in previous TTS models, it still suffers from slow generation speeds and the potential accumulation of errors due to its autoregressive nature. Recently, GAN models have been widely applied to vocoders, resulting in significant improvements in generation speed and the quality of generated audio. FastSpeech2 [14] and FastPitch [15] are acoustic models based on the Transformer [10] architecture. They generate mel-spectrograms non-autoregressively, resolving the problems of slow generation speeds and error accumulation encountered in the previous Tacotron model. FastSpeech2 incorporates a variance adaptor to learn various characteristics of speech, enabling control over pitch, speaking rate, and other attributes.

Voice synthesis models are evolving to generate natural and high-quality speech rapidly. Diffusion models are considered suitable for speech synthesis due to their ability to generate high-quality data. However, conventional diffusion models have slow generation speeds, making them unsuitable for real-time applications commonly used in TTS models. The demand for synthesis speed exists not only in the field of computer vision but also in speech synthesis, leading to ongoing research to improve the generation speed of diffusion models [16], [17]. One notable example is the denoising diffusion implicit model (DDIM) [16], which improves upon the structure of the denoising diffusion probabilistic model (DDPM) [5] to enhance generation speed. DDIM employs a non-Markovian approach instead of following the Markov chain used by original diffusion models. During the training phase, DDIM models the reverse process similar to DDPM, where the reverse process depends on increasing time steps. However, during the generation phase, it takes time steps as conditional inputs, allowing the model to skip intermediate stages from the initial noise $\mathbf{x}_T$ to directly reach an arbitrary time step $t$ state, denoted as $\mathbf{x}_t$, using accelerated sampling. This approach reduces the number of required time steps compared to DDPM. The denoising diffusion GAN model [17] applies the structure of GAN to diffusion models. The noise added or removed in the diffusion process follows a Gaussian distribution. In [17], the slow generation speed of diffusion models is attributed to the use of Gaussian distribution for sampling data during the reverse process. To address this issue, [17] introduces multimodal distributions for reverse diffusion, which accelerates the generation speed. Furthermore, to model these multimodal distributions, an adversarial training approach in GAN is employed, enabling the model to learn the distribution of noise required for the reverse process.

As the generation speed of diffusion models has been reduced, research is underway to apply diffusion models to acoustic models and vocoders in the field of voice synthesis. One notable example is Diff-TTS [18], which is an acoustic model utilizing the accelerated sampling technique used in DDIM. When generating speech without applying accelerated sampling, Diff-TTS demonstrates better audio quality compared to Tacotron2 or Glow-TTS [19]. On the other hand, when accelerated sampling is applied, the generation speed increases, but there is a trade-off with decreased audio quality. DiffGAN-TTS [20] is a state-of-the-art acoustic model based on the structure of denoising diffusion GAN. It supports voice synthesis for multiple speakers, and the generator is based on FastSpeech2, enabling control over the generated speech. DiffGAN-TTS may encounter limitations in effectively learning various elements in detail, as it requires a single discriminator to learn both the multimodal distribution for the reverse process and the characteristics of multiple speakers' voices.

In this paper, we propose an acoustic model called SpecDiff-GAN that combines diffusion model and GAN. This model incorporates the generator of DiffGAN-TTS [20] and is designed with two discriminators: a diffusion discriminator and a spectrogram discriminator. This design allows for the separation and independent training of features necessary for the reverse process and the characteristics of various speakers' voices. To evaluate the performance of the proposed model, we conduct experiments on multi-speaker speech synthesis. Objective evaluations are performed to assess how well the synthesized speech reflects speaker characteristics and the overall speech quality. The objective evaluations include measurements of structural similarity index measure (SSIM) [21], mel-cepstral distortion (MCD) [22], F0 root mean squared error (F0RMSE), short-time objective intelligibility (STOI) [23], perceptual evaluation of speech quality (PESQ)

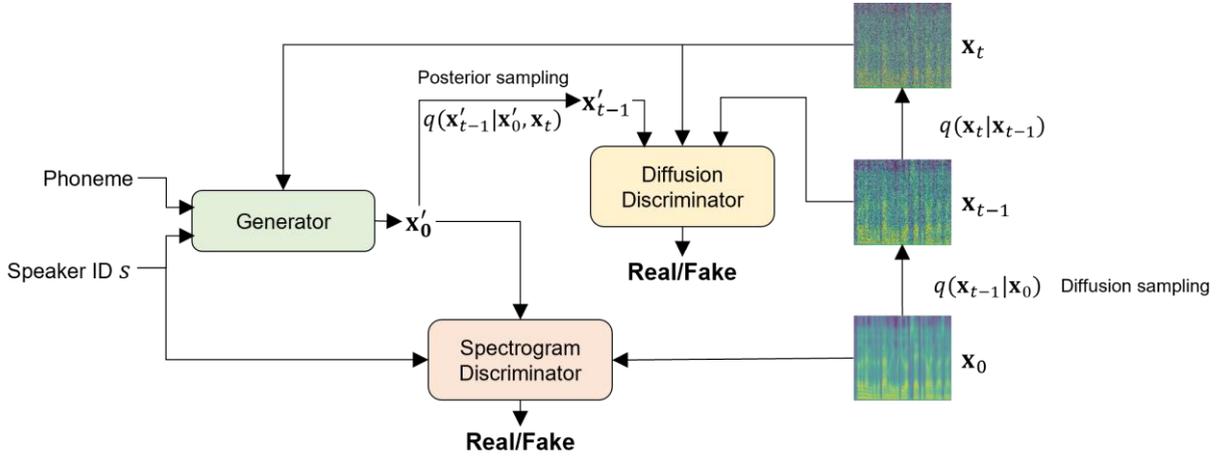

**FIGURE 1.** SpecDiffGAN-TTS architecture.

[24], and real-time factor (RTF). Additionally, subjective evaluations were conducted using comparative mean option score (CMOS) and similarity mean option score (SMOS).

The rest of this paper is organized as follows. In Section II, the operation principles of the diffusion model are explained, and Section III provides a description of the proposed model. In Sections IV and V, the experimental setup is described, and the experimental results of the proposed model are presented. Finally, in Section VI, the conclusions of the study are provided along with a discussion on future research directions.

## II. BACKGROUND

In TTS, diffusion models follow a *diffusion process* where Gaussian noise is gradually added to the (mel)-spectrogram through Markov chain transitions, converting it into a latent vector. They also employ a *reverse process* (also known as the *denoising process*) that removes noise from the latent vector and reconstructs the spectrogram. Let $\mathbf{x}_t \in \mathbb{R}^L$ for $t = 0, 1, \ldots, T$ be a sequence of corrupted spectrograms with the same dimension, where $t$ is the index for diffusion time step. Each diffusion process $q(\mathbf{x}_t|\mathbf{x}_{t-1})$ follows:

$$q(\mathbf{x}_t|\mathbf{x}_{t-1}) = \mathcal{N}(\mathbf{x}_t; \sqrt{1-\beta_t}\mathbf{x}_{t-1}, \beta_t \mathbf{I}), \quad (1)$$

where $\beta_t \in (0, 1)$ is a hyperparameter predefined ahead of model training. The whole diffusion process $q(\mathbf{x}_{1:T}|\mathbf{x}_0)$ is defined by Markov chain transition:

$$q(\mathbf{x}_{1:T}|\mathbf{x}_0) = \prod_{t \geq 1} q(\mathbf{x}_t|\mathbf{x}_{t-1}). \quad (2)$$

The reverse process is the inverse of the diffusion process and serves as the procedure for generating spectrogram from Gaussian noise. Each denoising process $p_\theta(\mathbf{x}_{t-1}|\mathbf{x}_t)$ follows:

$$p_\theta(\mathbf{x}_{t-1}|\mathbf{x}_t) = \mathcal{N}(\mathbf{x}_{t-1}; \mu_\theta(\mathbf{x}_t, t), \sigma_t^2 \mathbf{I}), \quad (3)$$

where $\mu_\theta(\mathbf{x}_t, t)$ and $\sigma_t^2$ are the mean and variance for the denoising model. The whole reverse process parameterized with $\theta$ is defined by:

$$p_\theta(\mathbf{x}_{0:T}) = p(\mathbf{x}_T) \prod_{t \geq 1} p_\theta(\mathbf{x}_{t-1}|\mathbf{x}_t). \quad (4)$$

Latent vector is gradually restored to a spectrogram through the reverse transitions $p_\theta(\mathbf{x}_{t-1}|\mathbf{x}_t)$. The goal of training is to maximize the likelihood $p_\theta(\mathbf{x}_0) = \int p_\theta(\mathbf{x}_{0:T}) \, d\mathbf{x}_{1:T}$, but since it is intractable, the model is trained to maximize the evidence lower bound (ELBO, $\mathcal{L} \leq \log p_\theta(\mathbf{x}_0)$) instead.

The key assumption in the diffusion model is that the noise levels at each step are small, which allows the diffusion process to be stable and the model to be tractable. The number of denoising steps $T$ is often assumed to be in the order of hundreds to thousands. Therefore, the parameters $\beta_1, \beta_2, \ldots, \beta_T$, which determine the amount of noise to be added during diffusion, are set to be small. As a result, the size of the diffusion time steps $T$ increases, leading to a significant time requirement for spectrogram reconstruction.

To reduce the number of denoising steps, Xiao *et al.* [17] proposed modeling the denoising distribution using a complex multimodal distribution. They introduced denoising diffusion GAN that model each denoising step using a multimodal conditional GAN. Their approach significantly reduced the denoising time steps to the extent that they became feasible for real-time applications. However, due to the limited number of time steps, the quality of the generated data may be lower compared to DDPM, and some features may not be adequately represented during the process of removing noise from the latent space and generating data.

## III. MODEL DEVELOPMENT

The overall architecture of the proposed SpecDiff-GAN in this paper consists of one generator and two discriminators, as shown in Figure 1. The DiffGAN-TTS architecture [20] serves

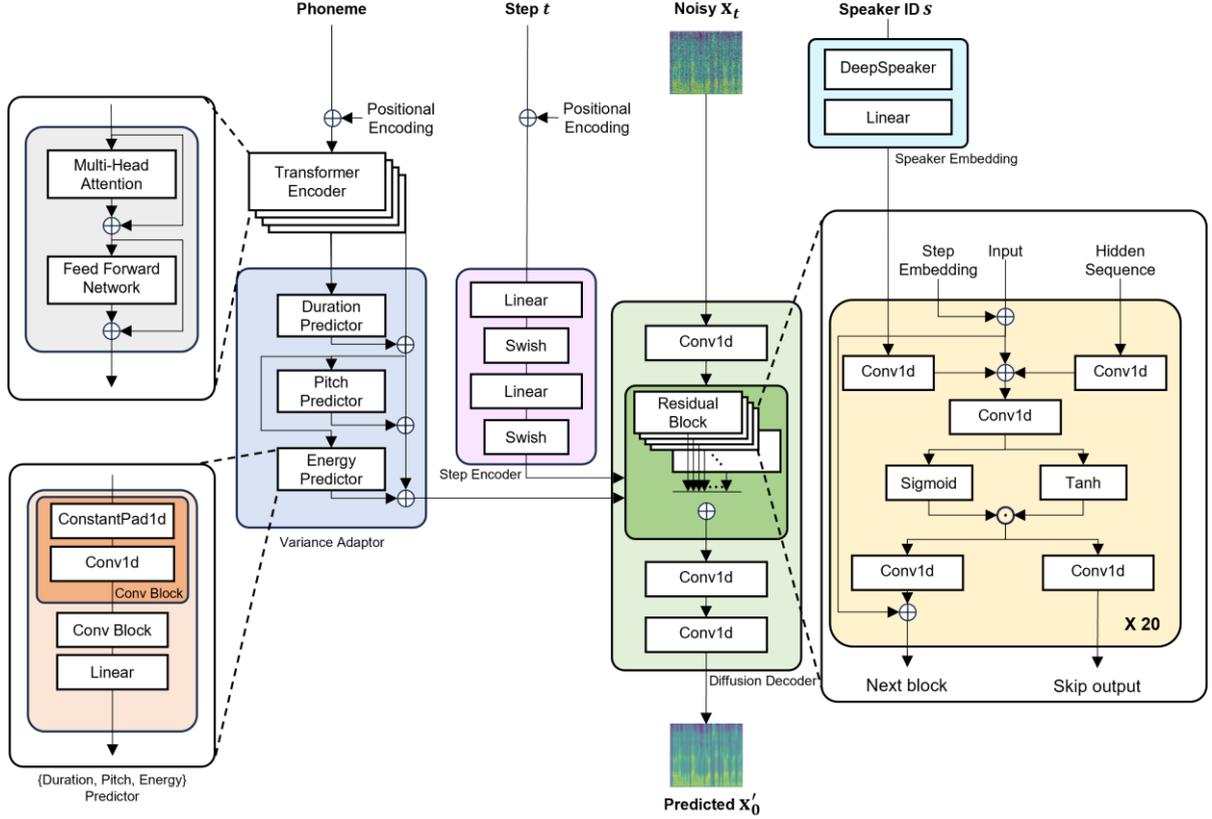

**FIGURE 2.** The generator structure of SpecDiff-GAN.

as the underlying base model for generating high-quality synthesized speech.

### A. THE GENERATOR OF SPECDIFF-GAN

The generator structure of SpecDiff-GAN is shown in Figure 2 and is based on the generator structure of DiffGAN-TTS. This generator is a modified version of the FastSpeech2 [14] architecture and consists of four transformer encoders, one variance adaptor, and one diffusion decoder. The transformer encoder takes the phoneme embedding sequence as input and converts it into a hidden sequence, similar to the FastSpeech2 approach. The variance adaptor is composed of a duration predictor, pitch predictor, and energy predictor. It observes the hidden sequence from the encoder to predict suitable speech length, pitch, and energy, thereby helping generate natural-sounding speech. The decoder is designed by replacing the decoder of the FastSpeech2 with a diffusion decoder.

The diffusion decoder in SpecDiff-GAN is based on the structure of WaveNet [13], an autoregressive vocoder model. However, there are differences between WaveNet and the diffusion decoder. While WaveNet vocoder uses dilated causal convolutions to generate waveforms, the diffusion decoder, as mel-spectrograms do not require as large receptive fields as waveforms, is implemented using conventional convolution layers. The proposed model's diffusion decoder takes speaker embeddings that reflect the characteristics of each speaker as conditional inputs, allowing it to generate different voices for different speakers. Deep Speaker [25] is used to obtain the speaker embeddings. Additionally, sinusoidal positional encoding [10] is applied to encode the time step $t$ at each time step, enabling the modeling of appropriate distributions for each time step.

The feature map generated through the transformer encoder and variance adaptor is passed through a $1 \times 1$ convolution layer and then inputted to each residual block of the diffusion decoder along with the speaker embedding. The hidden features outputted from each residual block are combined using skip connections and then passed through two $1 \times 1$ convolution layers to generate the mel-spectrogram.

In summary, the acoustic generator that produces the mel-spectrogram $\mathbf{x}'_0$ can be modeled as $G(\mathbf{x}_t, \mathbf{y}, s, t)$, where $\mathbf{x}_t$ is the corrupted mel-spectrogram, $\mathbf{y}$ is the phoneme input, $s$ is the speaker ID, and $t$ is the diffusion step index.

### B. THE DIFFUSION DISCRIMINATOR

The diffusion discriminator structure of the proposed model, as shown in Figure 3, is based on the discriminator structure of progressive growing of GANs (ProGAN) [26] with a time-dependent modification. The diffusion discriminator is composed of multiple downsampling convolution blocks. Each block consists of downsampling layers and 2D convolution layers. Timestep embeddings are conditionally applied to the input values to perform downsampling. The resulting extracted features from downsampling with conditioning and the extracted features from downsampling without conditioning are combined and passed as input to the next block. This process is repeated six times.

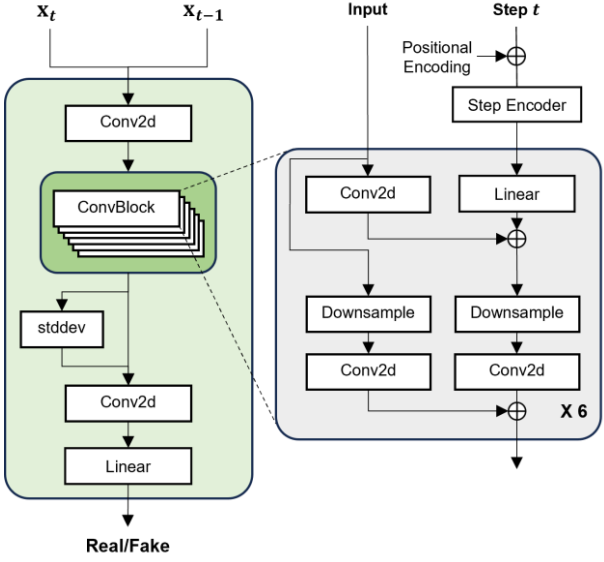

FIGURE 3. Diffusion discriminator of SpecDiff-GAN

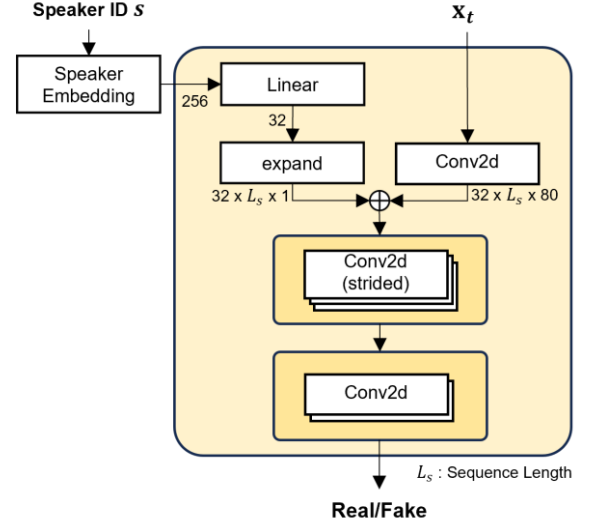

FIGURE 4. Spectrogram discriminator of SpecDiff-GAN

The discriminator takes $x_{t-1}$ and $x_t$ as inputs and iteratively compresses the data and extracts features to determine whether the reverse process from $x_t$ to $x_{t-1}$ is performed well. In ProGAN, to address the issue of mode collapsing that can often occur in GAN training, the minibatch discrimination technique [27] was employed, which calculates the closeness among the data within a batch. This technique is also applied in the diffusion discriminator to enhance the stability of the training process.

### C. THE SPECTROGRAM DISCRIMINATOR

The structure of the spectrogram discriminator in the proposed model is illustrated in Figure 4, taking inspiration from the structure of the multi-resolution spectrogram discriminator in UnivNet [28], a GAN-based vocoder model. While UnivNet utilizes multiple sets of short-time Fourier transform (STFT) parameters applied to mel-spectrograms for audio waveform generation, this paper focuses on using mel-spectrograms as an acoustic model for mel-spectrogram generation instead of waveforms. Therefore, the multi-resolution input is not used, and only the hierarchical structure of the discriminator in UnivNet is referred to.

In multi-speaker speech synthesis models, it is important to learn distinctive features for individual speakers. In this paper, speaker embeddings are utilized as conditional inputs to the discriminator. The speaker embedding goes through a linear layer, and the mel-spectrogram goes through a 2D convolution layer before performing broadcasting and addition operations. Afterward, the spectrogram discriminator is designed to repeat the convolution layers shown in Figure 4, allowing it to distinguish the characteristics of speaker-specific voices in the mel-spectrograms.

### D. LOSS FUNCTIONS

In this paper, the least squares GAN (LSGAN) [29] loss function is employed to train two discriminators. This loss function is utilized to prevent gradient vanishing and has already been proven to be effective when applied to the field of audio and speech synthesis. The diffusion discriminator, denoted as $D_d(x_{t-1}, x_t, t)$, is trained to minimize the loss:

$$\mathcal{L}_{diff} = \sum_{t=1}^{T} \mathbb{E}_{q(x_t)q(x_{t-1}|x_t)}[(D_d(x_{t-1}, x_t, t) - 1)^2] \\ + \mathbb{E}_{p_\theta(x_{t-1}|x_t)}\left[(D_d(x'_{t-1}, x_t, t))^2\right], \quad (5)$$

where $t$ denotes diffusion time step index. Equation (5) is the same loss function used in [20], but the speaker ID is not used as an argument. The spectrogram discriminator, denoted as $D_s(x_0, s)$, is trained to minimize the loss:

$$\mathcal{L}_{spec} = \mathbb{E}_{x_0 \sim p_{data}(x_0)}[(D_s(x_0, s) - 1)^2] \\ + \mathbb{E}_{x'_0 \sim p_\theta(x_{0:T})}\left[(D_s(x'_0, s))^2\right], \quad (6)$$

where $s$ denotes the speaker ID. The total loss of the SpecDiff-GAN discriminator is trained to minimize the following loss:

$$\mathcal{L}_D = \alpha \mathcal{L}_{diff} + (1-\alpha)\mathcal{L}_{spec} \quad (7)$$

where $\alpha$ is a mixing ratio hyperparameter that controls the resulting two losses.

In the SpecDiff-GAN generator, three loss functions are used. The feature matching loss is employed to ensure that the distribution of the generated data matches the distribution of the real data and to prevent the discriminator from overfitting. Feature matching loss $\mathcal{L}_{fm}$ is calculated by summing the $l_1$ distance between all discriminator feature maps of the generated and real data:

$$\mathcal{L}_{fm} = \alpha \mathbb{E}_{q(\mathbf{x}_t)} \left[ \sum_{i=1}^{N} \left\| D_d^i(\mathbf{x}_{t-1}, \mathbf{x}_t, t) - D_d^i(\mathbf{x'}_{t-1}, \mathbf{x}_t, t) \right\|_1 \right]$$
$$+ (1-\alpha) \mathbb{E}_{\mathbf{x}_0 \sim p_{data}(\mathbf{x}_0), \mathbf{x'}_0 \sim p_\theta(\mathbf{x}_{0:T})} \left[ \sum_{i=1}^{M} \left\| D_s^i(\mathbf{x}_0, s) \right. \right.$$
$$\left. \left. - D_s^i(\mathbf{x'}_0, s) \right\|_1 \right], \tag{8}$$

where $D_d^i(\cdot)$ represents the $i$-th hidden layer of the diffusion discriminator, and $D_s^i(\cdot)$ represents the $i$-th hidden layer of the spectrogram discriminator. $N$ and $M$ denote the number of hidden layers in the diffusion discriminator and spectrogram discriminator, respectively. The same mixing ratio $\alpha$ is used as set in the discriminator.

In addition to the feature matching loss $\mathcal{L}_{fm}$, the variance adaptor is trained using the acoustic reconstruction loss $\mathcal{L}_{recon}$ following [20] to accurately predict key characteristics of the speech, such as duration, pitch, and energy. Finally, based on [20], we train the generator to minimize the adversarial loss:

$$\mathcal{L}_{adv} = \sum_{t=1}^{T} \mathbb{E}_{q(\mathbf{x}_t)} \left[ \mathbb{E}_{p_\theta(\mathbf{x}_{t-1}|\mathbf{x}_t)} [(D_d(\mathbf{x}_{t-1}, \mathbf{x}_t, t) - 1)^2] \right]$$
$$+ \mathbb{E}_{\mathbf{x'}_0 \sim p_\theta(\mathbf{x}_{0:T})} [(D_s(\mathbf{x'}_0, s) - 1)^2]. \tag{9}$$

In total, the generator is trained by minimizing $\mathcal{L}_G = \mathcal{L}_{adv} + \mathcal{L}_{recon} + \lambda_{fm} \mathcal{L}_{fm}$, where $\lambda_{fm} = \mathcal{L}_{recon}/\mathcal{L}_{fm}$ following [30].

## IV. MODEL TRAINING AND EVALUATION

To validate the proposed SpecDiff-GAN model, we conduct multi-speaker speech synthesis experiments. We compare the proposed model with state-of-the-art alternatives to assess its ability to accurately capture speaker characteristics, generate smooth and natural pronunciation, and produce speech that closely resembles human-like sounds. The mel-spectrograms generated by the proposed SpecDiff-GAN are compared with both the ground-truth mel-spectrograms and those generated by FastSpeech2 [14] and DiffGAN-TTS [20]. To ensure a fair evaluation, all mel-spectrograms are converted into audio waveforms (i.e., speeches) using the HiFi-GAN vocoder [31].

### A. DATASET
The multi-speaker speech dataset used in this experiment is voice cloning toolkit (VCTK) corpus [32]. The VCTK corpus is an open dataset created by the University of Edinburgh in Scotland, which includes approximately 44 hours of English speech from 110 English-speaking speakers. Each speaker read 400 selected sentences from newspapers. All the included voices in the dataset were recorded in a non-reverberant indoor studio using high-performance microphones, with a sampling rate of 96 kHz and recorded in 24-bit audio. Subsequently, the audio was downsampled to a sampling rate of 48 kHz and provided as 16-bit audio. In the experiment, out of the total 44,000 voice samples available, 512 were used as the validation set, while the rest were used for training. To be used in the experiments, all the data was further downsampled to a sampling rate of 22.05 kHz.

**TABLE 1.** Considered hyperparameters for SpecDiff-GAN.

| Layer | Hyperparameters | Values |
|---|---|---|
| Generator | Number of diffusion time step | 4 |
| | Number of transformer encoder | 4 |
| | Number of attention head | 2 |
| | Number of residual blocks | 20 |
| Diffusion discriminator | Mixing ratio ($\alpha$) | 0.5 |
| | Number of convolution block | 6 |
| | Kernel size | {3} |
| Spectrogram discriminator | Number of Conv2d (strided) | 3 |
| | Number of Conv2d | 2 |
| | Kernel sizes | {3 × 3, 3 × 9} |
| | Stride height | 1 |
| | Stride widths | {1, 2} |
| | Padding height | 1 |
| | Padding widths | {1, 4} |

### B. MODEL SETUP AND TRAINING
The proposed SpecDiff-GAN model is implemented using the PyTorch framework. The *librosa* library is used to load audio files, the *audio* library is used to extract mel-spectrograms and energy from the audio, and the *parselmouth* library is used to extract fundament frequency F0. When converting audio to mel-spectrograms, the number of mel channels is set to 80, the hop size is set to 256, the window size is set to 1024, and the frequency range is set from 0 to 8000 Hz. The adam optimizer is used for training.

The hidden neuron count of one transformer encoder in the generator is 256, and with four encoders, it forms a feature space of 1024 dimensions. The encoder uses a convolution kernel size of 9. The variance adaptor consists of a duration predictor, pitch predictor, and energy predictor, each composed of two convolution blocks. The padding sizes for each predictor are (1, 1), (2, 2), and (3, 3), and the kernel sizes are (3, 3), (5, 5), and (5, 5) respectively.

In the spectrogram discriminator, zero padding height of size 1 and zero padding width of size 4 are applied before performing convolution with a 3 × 9 kernel. Only the second convolutional layer has a horizontal stride of 2 applied. The main hyperparameters used in SpecDiff-GAN are summarized in Table 1.

The workstation used for training has the Ubuntu 20.04 LTS operating system, and software dependencies are managed using Docker. The training of the proposed model utilizes four Nvidia A100 GPUs with 80 GB of memory. We utilized the publicly available codes [33] and [34] for FastSpeech2 and DiffGAN-TTS, respectively. All mel-spectrograms are converted to speeches using a pretrained HiFi-GAN vocoder. Our implementation and audio samples used for evaluation can be found on GitHub [35].

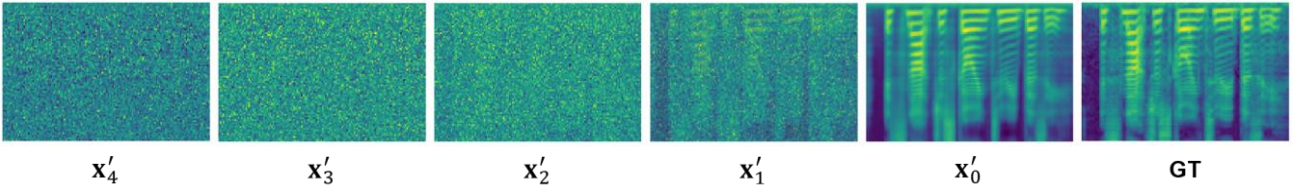

$\mathbf{x}'_4 \qquad \mathbf{x}'_3 \qquad \mathbf{x}'_2 \qquad \mathbf{x}'_1 \qquad \mathbf{x}'_0 \qquad$ GT

FIGURE 5. Visualization of the denoising process during inference of the proposed model, $\mathbf{x}'_0$ is the generated mel-spectrogram, GT is the ground-truth mel-spectrogram.

TABLE 2. Objective performance evaluation of speech synthesis. All spectrograms are converted to audio using a pretrained HiFi-GAN vocoder.

| Model | SSIM↑ | MCD↓ | F0 RMSE↓ | STOI↑ | PESQ↑ | RTF↓ |
|---|---|---|---|---|---|---|
| Mel-spectrogram (GT) + vocoder | - | **9.27** | - | 0.803 | 1.305 | - |
| FastSpeech2 [14] | 0.411 | 11.96 | 45.35 | 0.795 | 1.312 | **0.0035** |
| DiffGAN-TTS [20] | 0.787 | 9.50 | 33.48 | **0.813** | 1.295 | 0.0064 |
| SpecDiff-GAN (ours) | **0.791** | 9.35 | **33.19** | 0.805 | **1.326** | 0.0063 |

## C. PERFORMANCE METRICS

The performance evaluation metrics used are SSIM [21], MCD [22], F0 RMSE, STOI [23], PESQ [24], and RTF. While SSIM [21] is a metric used in computer vision to measure the similarity of images, in this paper, it is used to compare the similarity between the real mel-spectrogram and the generated mel-spectrogram by considering spectrograms as images.

MCD is a metric used to measure the difference between the mel-cepstral coefficients of real and generated speech, expressing it in [dB], which indicates the quality of the speech. MCD is given by:

$$MCD(\mathbf{x}, \hat{\mathbf{x}}) = \frac{10}{ln(10)} \sqrt{2 \sum_{i=1}^{N} \|\mathbf{x}_t - \hat{\mathbf{x}_t}\|_2}, \qquad (10)$$

where $\mathbf{x}$ and $\hat{\mathbf{x}}$ are mel-cepstrum of original and synthetic speeches, respectively, and $N$ is the total number of frames of speeches. A lower value indicates a higher similarity to the real speech.

F0 RMSE measures the difference between the ground-truth values and the generated values of the fundamental frequency F0. It is used to assess the accuracy of pitch in speech. A lower value indicates a better match between the ground-truth speech and the generated speech in terms of pitch. It is defined as:

$$F0\ RMSE = \sqrt{\frac{1}{N} \sum_{i=1}^{N} (F_i - \widehat{F_i})^2}, \qquad (11)$$

where $N$ is the total number of frames of speeches, $F_i$ is the F0 of the ground-truth speech, and $\widehat{F_i}$ is the F0 of the generated speech. A lower value indicates better performance in terms of F0 similarity between the ground-truth and generated speech.

STOI is a metric used to measure the intelligibility of speech. It is computed by comparing the short-term spectral characteristics and modulation information of the ground-truth speech and the generated speech. It provides a numerical value between 0 and 1, where higher values indicate better intelligibility. PESQ is a measure for assessing the quality of speech. It quantifies the perceptual similarity between a ground-truth speech and a generated speech. It measures various aspects of speech quality, including speech distortion, background noise, and overall perceptual degradation. RTF represents the ratio of the total processing time of the synthesized speech to the duration of the synthesized speech. If the RTF is less than 1.0, it indicates that the system can generate speech in real-time. A lower RTF value means that the system's speech generation speed is faster.

## V. RESULTS

The test set consists of voice recordings where English-speaking individuals spoke 512 sentences, one sentence each. Each voice sample has a duration ranging from approximately 1 second to 9 seconds. For comparison, DiffGAN-TTS and FastSpeech2 were also trained under the same conditions. The proposed model was trained for 300,000 steps, and the training took approximately 49.2 hours. To observe the process of generating data through the denoising process during inference, we visualized the output at each timestep in Figure 5. Through this figure, we can confirm that as the denoising process progresses, the generated output becomes more similar to the ground-truth mel-spectrogram.

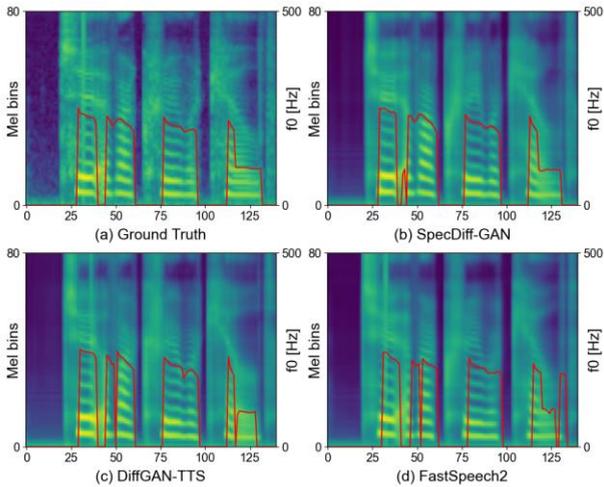

FIGURE 6. Comparison of mel-spectrogram with F0. (a) Ground Truth, (b) SpecDiff-GAN, (c) DiffGAN-TTS, (d) FastSpeech2.

TABLE 3. Subjective performance evaluation of speech synthesis. SMOS and CMOS evaluation. The CMOS score is the result of comparison with the proposed SpecDiff-GAN model. Values in ( ) are 95% confidence intervals.

| Model | SMOS | CMOS |
|---|---|---|
| Ground Truth (GT) | 4.44 (±0.07) | - |
| Mel-spectrogram (GT) + vocoder | **4.36** (±0.05) | 0.08 (±0.08) |
| FastSpeech2 [14] | 4.20 (±0.05) | -0.20 (±0.07) |
| DiffGAN-TTS [20] | 4.31 (±0.06) | -0.06 (±0.07) |
| SpecDiff-GAN (ours) | 4.35 (±0.06) | - |

## A. OBJECTIVE PERFORMANCE EVALUATION

Table 2 presents the evaluation of the performance of the models in generating mel-spectrograms. The time step for the diffusion model was set to 4 for all cases. As shown in Table 2, the proposed model demonstrates overall superior performance compared to the comparative models, especially in terms of MCD and F0 RMSE. These two metrics are indicators of comparing the characteristics and intonation of the speaker, making it evident that the spectrogram discriminator proposed in this paper effectively aids in generating mel-spectrograms that well learn the speaker and voice characteristics.

A high SSIM indicates that the mel-spectrograms generated by the proposed model closely resemble the ground-truth mel-spectrograms. Figure 6 displays the mel-spectrograms and fundamental frequency (F0) of both real and generated speech. While it may be difficult to visually perceive differences between the generated mel-spectrograms, the F0 of the proposed model is observed to closely approximate the ground truth compared to the F0 of the comparative models.

STOI shows lower performance compared to DiffGAN-TTS, which is likely due to the spectrogram discriminator incorporating some noise or reverberation when reflecting the characteristics of the speech, or the denoising process not entirely removing certain noise components. Regarding the RTF metric, FastSpeech2 achieved the highest measurement. The proposed model's RTF measured at 0.0063, indicating that it can generate speech much faster than the speed at which a human produces speech. This confirms that the proposed model can maintain a sufficiently fast generation speed while producing high-quality speech, making it suitable for real-time applications.

## B. SUBJECTIVE PERFORMANCE EVALUATION

Since all the mel-spectrograms were converted into speech using the same HiFi-GAN vocoder, it allows for a fair assessment of the quality of mel-spectrograms generated by the comparative models. Table 3 presents the evaluation results for SMOS and CMOS, comparing the proposed model with the comparison models. SMOS is a method used to evaluate the generated speech by presenting it to evaluators, who then rate its quality on a scale ranging from 1 (worst) to 5 (best) points. A score of 1 indicates that the generated speech is severely degraded and almost unintelligible, while a score of 5 signifies speech with no noise, free from awkwardness, and accurate pronunciation.

CMOS, with the proposed model as the reference, assigns a maximum of +3 points if the comparative model better represents the characteristics of the actual speech, and -3 points if it falls short in representing those characteristics. If there is no difference between the proposed model and the reference speech, 0 points are awarded. The original speech (i.e., ground truth) and the proposed model serving as the baseline are not subject to CMOS evaluation. The evaluation was conducted using a set of 30 speakers' voices, consisting of a total of 150 audio clips provided to the evaluators. These audio clips are unseen data, not utilized during the model training.

The SMOS scores indicate that actual recorded voice obtains the highest rating, followed by the reconstructed speech using the mel-spectrogram from the original speech through a vocoder. Among the models generating speech from text, the proposed model outperformed FastSpeech2 and DiffGAN-TTS, achieving higher scores. Therefore, it is observed that the proposed model generates more natural-sounding speech from text compared to the comparative models. This demonstrates that the spectrogram discriminator in the proposed model positively influences the speech quality and representation of speaker characteristics. The audio clips used for subjective evaluation can be found in [35].

TABLE 4. Ablation study results. Comparison of the effects of different component in SpecDiff-GAN discriminator.

| Model | SSIM↑ | MCD↓ | F0 RMSE↓ | STOI↑ | PESQ↑ |
|---|---|---|---|---|---|
| Baseline: SpecDiff-GAN (ours) | **0.791** | **9.35** | **33.19** | 0.805 | **1.326** |
| without spectrogram discriminator with speaker embedding | 0.779 | 11.50 | 40.58 | 0.798 | 1.296 |
| without spectrogram discriminator without speaker embedding | 0.789 | 11.63 | 40.03 | 0.804 | 1.323 |
| DiffGAN-TTS [20] (revisited) | 0.787 | 9.50 | 33.48 | **0.813** | 1.295 |
| FastSpeech2 [14] (revisited) | 0.411 | 11.96 | 45.35 | 0.795 | 1.312 |

### C. ABLATION STUDY

Table 4 presents the results of an ablation study conducted to examine the impact of the spectrogram discriminator on the model's performance. Two experiments were conducted: one involved completely removing the spectrogram discriminator, and the other involved removing the spectrogram discriminator while connecting the speaker embedding, which was connected to the spectrogram discriminator, to the diffusion discriminator.

The experimental results confirmed that both models without the spectrogram discriminator exhibited a decrease in performance. Particularly, the model with the speaker embedding connected to the diffusion discriminator experienced even greater performance degradation. This observation suggests that learning both speaker information and multimodal distributions for the reverse process in a single discriminator adversely affects the model's performance. These results confirm that the presence of the spectrogram discriminator contributes to the improvement of the model's performance.

## VI. CONCLUSIONS

In this paper, we propose the SpecDiff-GAN model, which combines a diffusion model and GAN to enhance the quality of generated speech while maintaining the generation speed. The proposed model features a dual discriminator structure, comprising the diffusion discriminator and the spectrogram discriminator. The diffusion discriminator models the multimodal distribution for the reverse process, thereby accelerating the generation process of the diffusion model. On the other hand, the spectrogram discriminator is designed to differentiate between the original and generated mel-spectrograms, enabling the generator to produce high-quality speech that closely resembles the original. To evaluate the model's performance, we utilized five objective evaluation metrics and two subjective evaluation metrics. Comparative analysis with FastSpeech2 and DiffGAN-TTS demonstrated that the proposed SpecDiff-GAN model outperforms the comparative models. Additionally, the ablation study confirms the effectiveness of our proposed techniques in improving the model's performance. The use of the dual discriminator shows promising results in performance improvement, and we are conducting further research to enhance the discriminator's performance.